\documentclass{appolb}
\usepackage{graphicx}
\usepackage[sorting=none, backend=biber,maxnames=10]{biblatex}
\usepackage[utf8]{inputenc}    % utf8 support       %!!!!!!!!!!!!!!!!!!!!

\newcommand{\oPs}{\mathrm{oPs}}

\begin{document}
\title{ Feasibility of ortho-positronium lifetime studies with the J-PET detector in context of mirror matter models.%
}
\author{Wojciech Krzemien
\address{High Energy Physics Division, National Centre for Nuclear Research, \\
 05-400 Otwock-\'Swierk, Poland \\
 wojciech.krzemien@ncbj.gov.pl}
 \and
Elena P{\'e}rez del R{\'i}o
\address{LNF-INFN, Laboratori Nazionali di Frascati,\\
Via E. Fermi 40, 00044, Frascati (RM), Italy\\
eperez@lnf.infn.it}
 \and
 Krzysztof Kacprzak
 \address{Faculty of Physics, Astronomy and Applied Computer Science \\Jagiellonian University\\%Nazionali di Frascati,\\
S. \L{}ojasiewicza 11, 30-348 Krak\'ow, Poland \\%(RM), Italy\\
k.kacprzak@alumni.uj.edu.pl}
}

\maketitle
\begin{abstract}
We discuss the possibility to perform the experimental searches of invisible decays
in the ortho-positronium system
with the J-PET detector. 
%
%\keywords{Positronium; fundamental symmetries; invisible decays; CP violation}
\end{abstract}
%\PACS{12.20.Ds, 12.20.Fv, 14.70.Pw, 36.10.Dr}

\section{Introduction}
In spite of the remarkable success of the Standard Model (SM) of particles, which precise predictions have been confirmed by many experiments, SM fails to explain all features of Nature and still several important questions remain without answers. One of such open puzzles is the existence and the nature of the so-called  Dark Matter (DM) which is believed to be described by new particles. 

DM is called \textit{dark} because it does not interact with electromagnetic forces, which makes its experimental observation extremely difficult. 
However, since DM particles have invariant mass, they can act on the ordinary matter via the gravitational force. Indeed, several astrophysical observations
have been interpreted as indirect evidence of the DM existence~\cite{ext1, ext2, ext3, ext4, ext5}.

In laboratories, many attempts have been performed to detect the DM particles~\cite{lhcb, babar, na48, kloe}. However, so far neither direct nor indirect experimental evidence has been found.

In mirror matter (MM) models, the mirror (spatial parity-) counter-partners for the ordinary SM particles are introduced. Since such objects would interact with SM matter mainly via gravitational force, the mirror particles seem to be natural DM candidates.

The search for mirror matter particles can be carried out in the so-called invisible decays of positronium (Ps). 
Ps is a bound state of electron and positron. 
Its ground state has two possible configurations, a singlet state $^1S_0$, para-positronium (pPs) where the spins of the electron and the positron are anti-parallel, and the $^3S_1$ triplet, ortho-positronium (oPs), where the spins are parallel.
Being a purely leptonic system, Ps is precisely described by Quantum Electrodynamics (QED) with very small radiative corrections from Quantum Chromodynamics (QCD) and weak interaction effects. Those properties make it an attractive system for various experimental tests~\cite{steven1,moskal:sym}.
Assuming MM model, oPs can oscillate into its parity-partner mirror oPs$^{\prime}$, which consequently decays into mirror photons not detectable in the laboratory. Experimentally, this process would increase the observed oPs decay rate.

In this contribution we discuss the feasibility of the measurement for the mirror matter in the invisible decays with the J-PET device.
The article is structured in the following way:
In section~\ref{sec:mirror} we shortly discuss the mirror oscillations phenomena in the context of Ps invisible decays.  
In section~\ref{sec:exp} the experimental methods are briefly described. 
Next, in section~\ref{sec:jpet} the J-PET detector is introduced. The experimental prospects are discussed in~\ref{sec:prospects}.
Finally, section~\ref{sec:outlook} contains the outlook.

\section{Mirror matter and invisible decays }
\label{sec:mirror}

The concept of MM was first introduced by Lee and Yang~\cite{Lee:1956qn} in the famous paper in which they proposed a series of experiments that led to the confirmation of spatial parity violation in weak interactions~\cite{Wu,Garwin}. To restore the spatial parity symmetry in a larger sense, Lee and Yang suggested the existence of parity reflected counter-partners of the ordinary matter.

This idea was further developed into MM models~\cite{Salam:1957,Okun:1966}, in which the hidden mirror sector contains not only  particles but also interactions. The mirror world would consist of the parity reflected states (e.g. right-handed for mirror particles and left-handed for antiparticles) of every ordinary particle from SM.

Apart from the gravity,  MM could interact with ordinary matter via a mixing mechanism, as proposed by Glashow~\cite{glashow}, to explain the experimental results of oPs decays. 
In this model the invisible decay of oPs can proceed through the annihilation into a virtual single-photon, which oscillates into a mirror photon, connecting oPs and the corresponding mirror oPs$^{\prime}$ partner
via the kinetic term :
\begin{equation}
    \label{eq:dm_lagrangian}
    \mathcal{L} = \epsilon F^{\mu\nu}F^{'}_{\mu\nu}
\end{equation}
where $\epsilon$, $F^{\mu\nu}$ and $F^{'}_{\mu\nu}$ are the mixing parameter, and field tensors for electromagnetism and mirror electromagnetism respectively~\cite{glashow}.

We can express the probability that a oPs does not change its states in a time t, assuming that it was in the oPs state at t=0, as: 
\begin{equation}
P(t) = e^{-\Gamma t}\times \cos^2(2 \pi \epsilon f t), 
\label{eq:oscill}
\end{equation}
where $\Gamma$ and $f$ are the decay rate of oPs into three photons, and the contribution to the ortho-para splitting from the one photon annihilation diagram involving oPs ($f = 8.7 \times 10^4$ MHz) respectively.
The inclusion of oscillation process would lead to the increase of the measured lifetime. 
The complementary probability that oPs oscillates into the mirror state in the time \textit{t} is given by: 
\begin{equation}
P'(t) = e^{-\Gamma t} \times \sin^2(2 \pi \epsilon f t), 
\end{equation}

The branching ratio $Br (\oPs \to {\rm invisible})$ can be expressed as an average probability over a long time of measurement
\begin{equation}
    Br (\oPs \to {\rm invisible})
    =\Gamma \times \int_{0}^{\infty} P'(t) dt= \frac{2 (2 \pi \epsilon f )^2}
    {\Gamma^2 + 4 (2 \pi \epsilon f )^2}.
\end{equation}
    
%integral_0^∞ exp(-A t) sin^2(B t) dt = (2 B^2)/(A^3 + 4 A B^2) for A>0
Values of $\epsilon$ between $\sim 10^{-10}$ and $\sim 4 \times 10^{-9}$, constraint by DM models aiming to explain the DAMA anomaly \cite{Foot:2019}, correspond to branching ratio expectations between $5 \times 10^{-10}$ and $2 \times 10^{-7}$. These calculations do not consider incoherent processes (e.g. collisions with matter), and thus apply only to oPs decays in vacuum.
Another constraint can be determined from the prediction of the primordial $^4$He abundance by the SM and it gives $\epsilon \leq 3 \times 10^{-8}$~\cite{Carlson:1987si}.

In principle, the experimental confirmation of the invisible oPs decays can be interpreted not only in the frame of the MM models but also in context of other  
new physics scenarios, e.g. milli-charged particles and extra space-time dimensions \cite{Gninenko:2006sz}. 

\section{Experimental methods}
\label{sec:exp}
The search for invisible decays is a major focus 
for Ps decay experiments
\cite{Atoian:1989tz, Mitsui:1993ha, Gninenko:1994dr, Badertscher:2006fm, vigo}.The process allowed by SM that would mimic the new physics invisible decays has a experimentally negligible branching ratio $Br(\mathrm{o}$-$\mathrm{Ps} \rightarrow \nu_{e} \bar{\nu_{e}}) < 10^{-18} $~\cite{CZarneckiOR}.

So far, two types of experimental methods have been used.
The first one is based on the precise measurement of the oPs decay rate looking for a difference  relative to the QED predictions due to some unconventional processes, e.g. due to the oscillation of the oPs into the mirror world.
The second, a direct search approach, relies on the definition of a \textit{no-signal} region in some experimental distribution, e.g. the sum of the energies deposited by the oPs decay products, which corresponds to detector response not connected with any physical processes but, e.g. with detector noise or other nuisance phenomena. The observation of any excess in the \textit{no-signal} region could be interpreted as a manifestation of the invisible decay mode.

The \textit{no-signal} method was used in two independent measurements by the ETH Z\"urich group. They established the most accurate constraint on a possible invisible decay of oPs in vacuum \cite{vigo} up to now\footnote{There is a new result from the ETHZ group in e-print~\cite{vigo2} updating the BR measurement to $\mathcal{O}(10^{-5})$}:
\begin{equation}
    \mathrm{BR}(\mathrm{o \!-\! Ps} \rightarrow invisible) < 5.0 \times 10^{-4}, ~~~~ 90\% C.L.
\end{equation}
which can interpreted as a constraint on the mixing parameter: $\epsilon < 3.1 \times 10^{-7} $.

The most accurate measurements of the oPs decay rate are consistent with each other and with the theoretical prediction known up to two-loop ($\mathcal{O}(\alpha^{2})$) corrections (see Eq.1 in ~\cite{steven1}).

The Tokyo group ~\cite{kataoka} obtained:
\begin{equation}
    \Gamma = 7.0401 \pm 0.0007 \times 10^{6}~s^{-1}
\end{equation}
with the oPs produced in $\mathrm{SiO}_{2}$ powder, whereas Ann Arbor group~\cite{vallery} measured:
\begin{equation}
    \Gamma = 7.0404 \pm 0.0010 \pm 0.0008 \times 10^{6}~s^{-1}
\end{equation}
with a slow positron beam on silica target. These results are consistent with the QED theory predictions, with the caveat that the present experimental uncertainties on the decay rate are about 100 times larger than the theoretical error, and thus the sensitivity needed to test the oPs mirror component.

In experiments, oPs is typically formed not in the vacuum, but in some kind of material where it can interact with the environment. Because the interaction of mirror oPs$^{\prime}$ with the standard matter is negligible, it follows that one has to consider interaction of just the normal oPs with the host material. The net oscillation effect can be suppressed by these processes, e.g. scattering with the gas molecules or with the cavity walls~\cite{Demidov:2012}. These effects must be taken into consideration when performing the experiment because they effectively reduce the experimental sensitivity for extracting the mixing term.

\section{J-PET detector}
\label{sec:jpet}
\
 The J-PET (Jagiellonian-Positron Emission Tomography scanner) is a high acceptance multi-purpose detector optimized for the detection of photons from positron-electron annihilation.
 Originally designed as a medical scanner, the J-PET detector has several advantages such as very good timing resolution~\cite{timing}, possibility of data taking in the continuous mode (triggerless)~\cite{Korcyl-Acta,Korcyl-IEEE}, fully digital front-end electronics~\cite{j-pet-jinst},  and efficient discrimination between different Ps decay channels~\cite{Daria:epj2016}. These make it suitable in a broad scope of interdisciplinary investigation, e.g. medical imaging~\cite{moskal:pmb2019, NATURE}, fundamental symmetry tests~\cite{moskal:sym} and quantum entanglement studies with oPs~\cite{ENTANGLEMENT}.
 
The J-PET  device  is made of plastic scintillators. The current prototype is built from three cylindrical layers (radius of 42.5, length of 50 cm). Light signals from each strip are converted to electrical signals by photomultipliers placed at opposite ends of the strip~\cite{NIM2015, J1, Raczynski2014}. 

Presently, a new innermost layer is being installed and commissioned, with the start of data acquisition of the full 4 layer setup planned for the 2019 winter~\cite{moskal:pmb2019}. 
\begin{figure}[htb]
\centerline{%
\includegraphics[width=7cm]{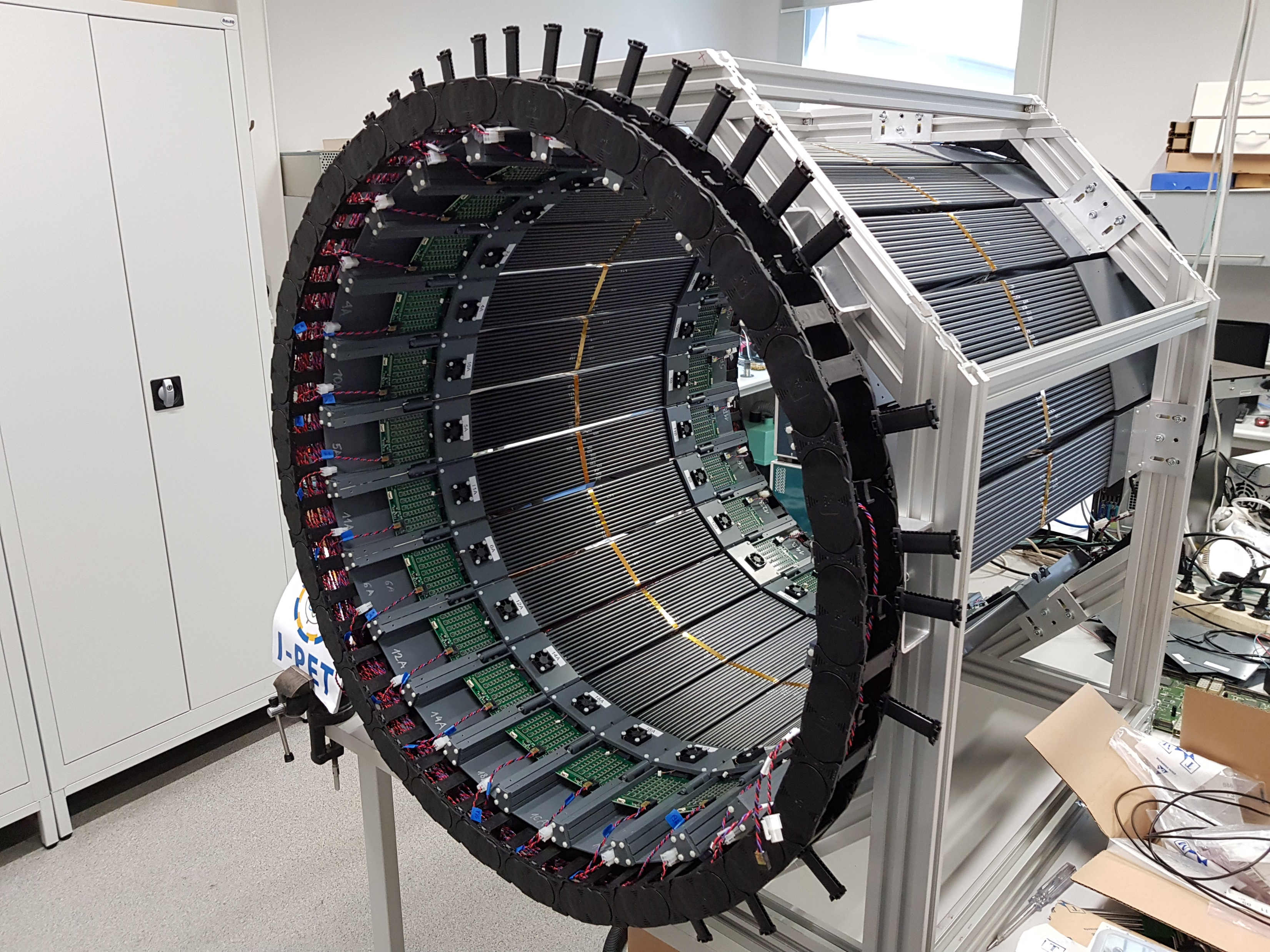}
\hspace{0.5cm}
\includegraphics[width=6cm]{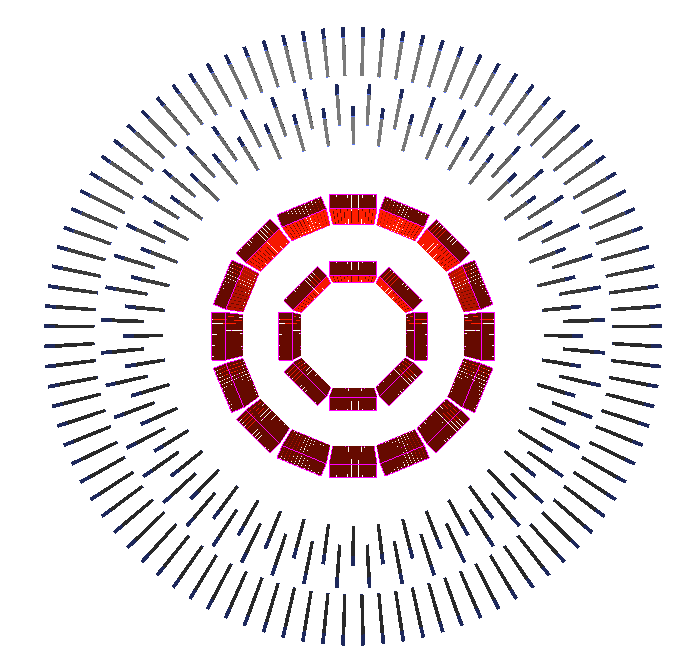}
}
\caption{(Left) The fourth layer of the J-PET detector. (Right) J-PET detector scheme with modules from the 4th layer rearranged into 2 internal layers.
\label{fig:acc}
}
\end{figure}
This fourth layer, see Fig.~\ref{fig:eff} (Left) is read out by matrices of silicon photomultipliers (SiPM), which is expected to triple the efficiency for the single photon detection and improve the time resolution by about a factor of $1.5$~\cite{timing}.
In addition, the new setup allows to recombine the modules into various arrangements that can be adapted for dedicated physics measurements. E.g. the setup in Fig.~\ref{fig:acc} (Right), contains two modular layers. Future measurements will be carried out with the fully equipped detector. 
 
\section{J-PET prospects}
\label{sec:prospects}

The J-PET detector is suited to perform  the precise measurement of oPs lifetime using the decay process oPs $\rightarrow 3 \gamma$. The results will be confronted with high precision QED calculations. Any significant discrepancy between theoretical predictions and the measurement would point into the direction of new physics and might be interpreted in the frame of MM models. 

In the J-PET experiments, a $^{22}\mathrm{Na}$ source can be used. The $\mathrm{Na}$ isotope decays through $\beta^+$ transitions, emitting a positron that slows down in its interactions with matter, reaching thermal energies. Then it undergoes free annihilation or forms positronium.

The Ps is produced mostly in the ground state, forming pPs or oPs with probabilities of 25~\% and 75~\%, respectively. However, the interactions with the surrounding matter can lead oPs to spin inversion or to pick-off processes and, as a result, significantly decrease the relative ratio of $3\gamma/2\gamma$ annihilation. 
In the J-PET setup, an amberlite porous polymer XAD-4 (CAS 37380-42-0) material placed in a vacuum chamber increases the effective yield of annihilation into $3\gamma$ to 29\%~\cite{Jasinska-Acta-XAD4}. 

The $\beta^+$ decay from $\mathrm{Na}$ source is followed - after few picoseconds - by the emission of a monochromatic photon of 1.27 MeV ($\gamma_{de-ex}$) from the de-excitation of neon nucleus. The de-excitation gamma registration time enables to set a window for the registration of the annihilation photons and for lifetime measurements.

\begin{figure}[htb]
\centerline{%
%\hspace{0.5cm}
\includegraphics[width=4.5cm]{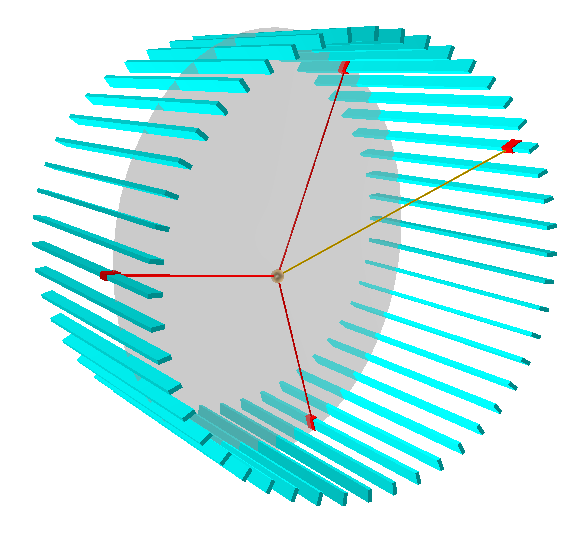}
%\hspace{0.5cm}
\includegraphics[width=4.5cm]{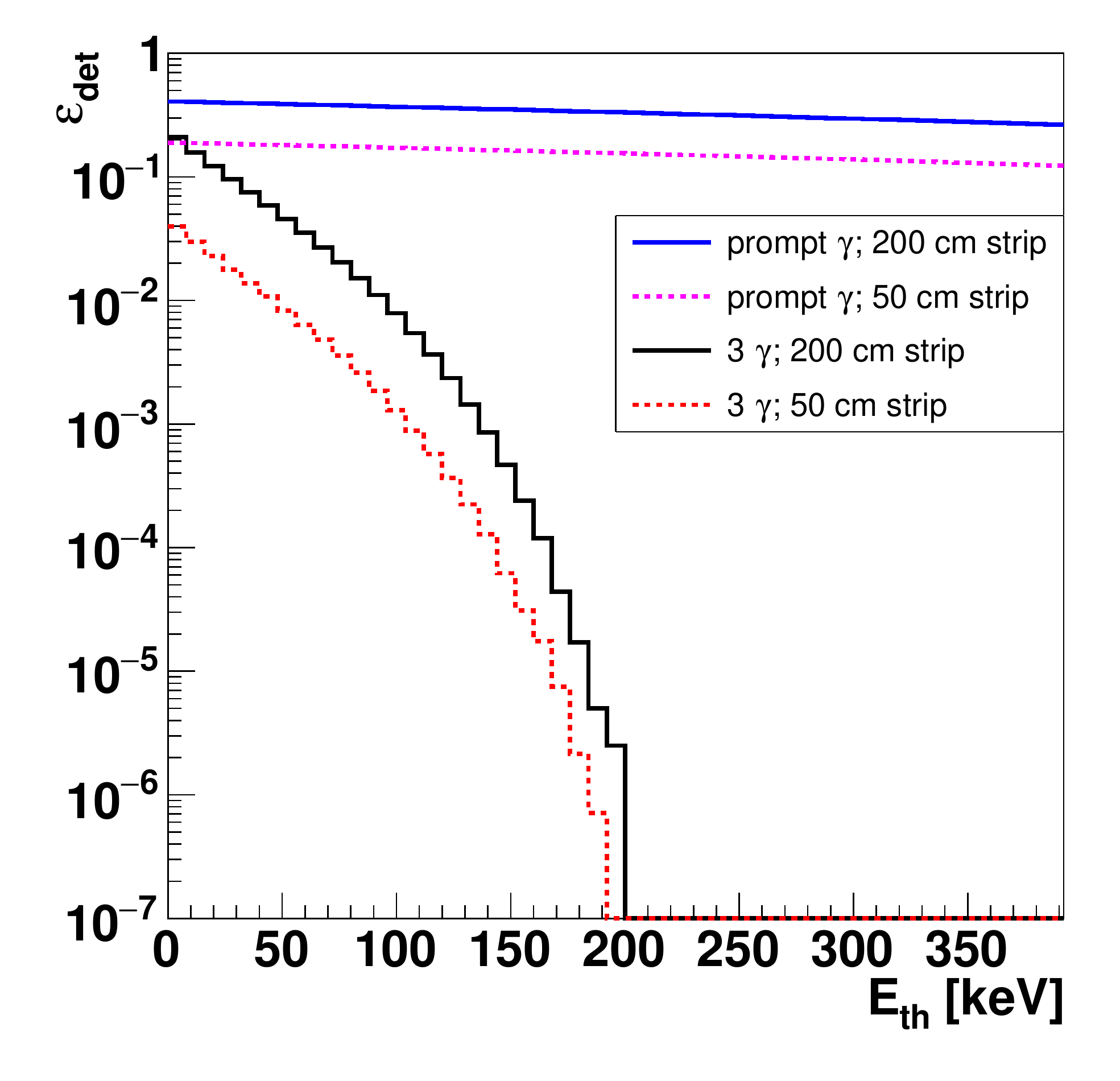}
%\hspace{0.5cm}
\includegraphics[width=4.5cm]{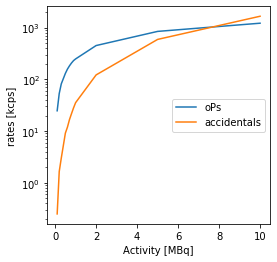}
}
\caption{
(Left) Inner thin layer of the J-PET detector with superimposed decay plane of 
$o\!-\!Ps \to  3\gamma$ process. 
Lines inside the plane indicate annihilation photons and the  
line pointing out the $\gamma_{de-ex}$.  
(Center) Registration efficiency for $\gamma_{de-ex}$ (upper lines) and $\mathrm{o}$-$\mathrm{Ps} \rightarrow 3\gamma$ as a function of the energy-loss threshold, estimated for a single-layer setup.  The dashed and solid lines indicate efficiencies for
50 cm and 200 cm strip length, respectively~\cite{moskal:pmb2019}.
(Right) Estimated rates of produced accidental and oPs events as a function of the source activity. The registration efficiency is not included in the estimate.
\label{fig:eff}
}
\end{figure}

The events can be efficiently reconstructed and separated using the angular and timing resolution of the J-PET detector. Subsequent constraints in hit position, energy and time allow to disentangle the de-excitation from the annihilation photons, at the same time that $3\gamma$ and $2\gamma$ events can be selected. Also, due to the kinematic constraints of the decay, the vertex can be precisely reconstructed using the trilateration technique~\cite{alek:pra2016}, see Fig.~\ref{fig:eff} (Center).

In order to reduce the main source of background, consisting of pick-off events, a selection based on two-photon events can be performed in the same data sample. This would allow to evaluate and subtract the background component directly from data. 

To better distinguish between the multi-photon decays categories we plan to use Machine Learning (ML) algorithms, such as Boosted Decision Trees (BDT) or Artificial Neural Networks (NN). This way, the sample can be used to train on two photon events in order to properly separate the pick-off from the interesting three-photon decays. A similar method has been already proposed by the ETHZ group~\cite{crivelli:2018}, showing a good separation between the two categories of events.

The level of the statistical uncertainty that can be achieved  is inversely proportional to the number of generated Ps multiplied by the registration efficiency for the given process. The BR(o-PS $\rightarrow X$) can be expressed as $\mathrm{BR} \approx \frac{1}{N_{Ps} \times \epsilon_X}$.

The advantage of the short signals formed in the plastic scintillators (of the order of few ns) allows to significantly reduce the pile-up problems present in the previous experiments~\cite{vigo,vallery,asai:2003} and therefore to choose sources with higher activities.
The estimate of the accidental coincidence rate together with the oPs production rate is shown in Fig.~\ref{fig:acc} (Right). The accidental event is defined as an event in which more than oPs decays take place in the 200 ns frame calculated with respect to the time of the de-excitation gamma. For the further estimates we use a 1 MBq source to keep the signal-to-background ratio reasonable for the measurement.

With this assumption in one years of data taking with J-PET detector we expect to have about ${N_{oPs}\approx 10^{13}}$ oPs, and about the same amount of pPs produced.

As aforementioned, the current best experimental accuracy of lifetime measurement $ 10^{-4}$ is 100 times less precise than the theoretical calculations~\cite{steven1}.
To reach the statistical uncertainty below  $10^{-4}$ one needs to reconstruct $10^{9}$ oPs events. For this purpose the new modular J-PET needs to be used in a double-layered configuration, see Fig.~\ref{fig:acc} (Left). This arrangement corresponds  roughly to an increase of detection efficiency for single photon of about a factor of 2. Taking into account the estimated number of oPs generated using the J-PET setup ($\approx 10^{13}$) and the efficiency for the detection of annihilation photons (2\%) and of the de-excitation photon (about 20\%) see Fig.~\ref{fig:eff} (Right) together with the double-layered configuration, a sensitivity below $10^{-5}$ could be reached after two years of data taking using the J-PET detector.

\section{Summary and outlook}
\label{sec:outlook}
In this article, we discussed the capability of the J-PET detector to perform searches of oPs invisible decays with the precise measurement of oPs lifetime distribution.
After two years of data taking the expected sensitivity would allow for more precise tests of the QED predictions. 
Any possible discrepancy between the measured and predicted oPs decay rate could be interpreted in terms of invisible decays, with possible MM model interpretation.
The estimated statistical sensitivity is rather conservative and can be largely improved by adding a new set of layers, which would further increase the detection efficiency. 
Currently, we carry out the studies of the ML-based method for the pick-off background subtraction.  

Another possibility to decrease the background and increase the experimental reach of the measurement is the implementation of a tagger system for the positrons emitted by the source, based on SiPM and scintillator fibers.
Finally, the proposed search can be extended to the measurement carried out not only with radioactive sources but with a positron beam. The new version of the modular J-PET detector is highly portable and would allow transporting the device to facilities with positron beams such as e.g.  Trento Institute of Fundamental Physics and Applications, in Italy or ETH Z\"urich in Switzerland, to perform new and independent measurements.

\section*{Acknowledgments}
We would like to  thank S.~Bass, P.~Crivelli, and P.~Moskal,  for stimulating discussions. We would like to thank S.~Sharma for sharing
the 4th layer figure.

This work was supported in part 
by the Foundation for Polish Science through the Grant No.\
TEAM POIR.04.04.00-00-4204/17.
\printbibliography
\end{document}